\documentclass[epj]{svjour}
\usepackage{latexsym}
\usepackage{graphics}
\usepackage{epsfig}
\begin{document}
\title{Antideuteron production in proton-proton and proton-nucleus
collisions}
\author{R.P. Duperray\inst{1}\thanks{duperray@isn.in2p3.fr},
 K.V. Protasov\inst{1}\thanks{protasov@isn.in2p3.fr},
 \and A.Yu. Voronin\inst{2}\thanks{avoronin@aha.ru}
}                     % Do not remove

\institute{Institut des Sciences Nucl\'eaires, IN2P3-CNRS, UJFG,
53, Avenue des Martyrs, F-38026 Grenoble Cedex, France \and
Lebedev Physical Institute, Leninsky prospekt, 53, 117924 Moscow,
Russia }
\date{Received: date / Revised version: date}

\abstract{ The experimental data of the antideuteron production in
proton-proton and proton-nucleus collisions are analyzed within a
simple model based on the diagrammatic approach to the coalescence
model. This model is shown to be able to reproduce most of
existing data without any additional parameter.  } %end of abstract
\maketitle

\section{Introduction}

The interest in the study of production of light antinuclei, in
particular, antideuterons in proton-proton and proton-nucleus
collisions has recently intensified. There are, at least, two
major reasons for this. Firstly, studies of antideuteron
production in cosmic space can be a very powerful tool to search
for antimatter in the Universe \cite{Salati}. It is planed to
measure the antideuteron flux in the future AMS
\cite{AMS} and PAMELA \cite{PAMELA} experiments.

Secondly, the possibility to make experiments with antideuteron
beams was discussed recently (see \cite{Iazzi99} and references
therein).

The theoretical estimation of antideuteron production preformed in
the article \cite{Salati} are based on the well-known coalescence
model \cite{CoalModel} which supposes that two nucleons fuse into
a deuteron if the momentum of their relative motion is smaller
than a certain quantity $p_0$, the coalescence radius in momentum
space. This momentum $p_0$ is considered as a free parameter to be
fixed from the experimental data.

More than ten years ago, a quite simple diagrammatic approach to
the coalescence model provided a microscopical basis for the
coalescence model and expressed the parameter $p_0$ in terms of
the slope parameter of the inclusive nucleon spectrum and the wave
function of the produced nucleus \cite{Kolyb}. Within this
approach, it appears to be possible to explain the empirical fact
of approximate equality of the values of the coalescence radii for
the description of the yields of various light fragments under
similar kinematical conditions.

The aims of this article are to generalize this diagrammatic
approach to antinuclei production (by introduction of threshold
effects and by taking account of anisotropy of angular
distributions) and to apply this model to the antideuteron
production in proton-proton and proton-nucleus collisions.
Particular attention will be paid to proton-proton collisions due
to their interest for astrophysics. We will show that, in the
cases where the inclusive antiproton production cross-section and
deuteron wave function are well-known, this approach can describe
quite well the inclusive antideuteron production cross-section
without any additional parameter.

Note, that there are few articles in which the anti\-deuteron
production is discussed within different approaches
\cite{Mro,Dover,LeuHeinz,SchHeinz}. Some of them require
additional parameters to describe the experimental data and none
of them describes the whole ensemble of experimental data.

The article is organized as follows. In section 2 the main ideas
of the diagrammatic approach to the coalescence model are
described and this approach is generalized to the case of
antideuteron production. In Section 3, the description of the
experimental data are presented. Finally, we provide a brief
summary of the results.

%%%%%%%%%%%%%%%%%%%%%%%%%%%%%%%%%%%%%%%%%%%%%%%%%%%%%%%

\section{Diagrammatic approach to the coalescence model}

Let us remind the reader the main ideas of the diagrammatic
approach to the coalescence model \cite{Kolyb}. As a basis for the
coalescence model, the simplest Feynman diagram of Fig. 1,
corresponding to fusion of two nucleons is considered. Here the
symbol $f$ denotes the state of all other particles except the
nucleons 1 and 2 which form the deuteron. The physical picture
behind this diagram is quite simple: the nucleons produced in a
collision (block A) are slightly virtual and can fuse without
further interaction with the nuclear field. This simplest diagram
is not the only possible contribution. However, as was shown in
\cite{Braun}, there are mutual cancellations of a number of other
diagrams and, as a result, the diagram of Fig. 1 is the dominant
one.

%---------------------------------------------------------------------------
\begin{figure}[h!]
\epsfxsize=8cm \centerline{\epsfbox{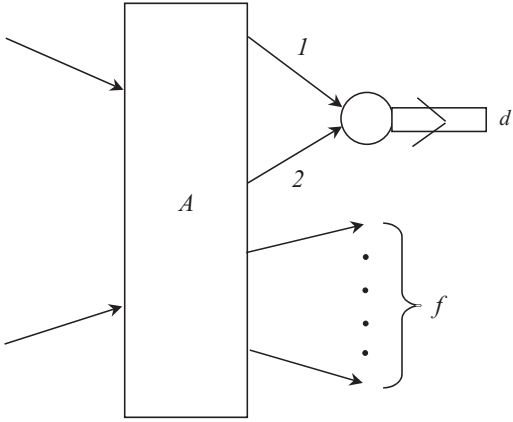}} \caption{ The
simplest Feynman diagram corresponding to coalescence of two
nucleons into a deuteron. } \end{figure}
%
%---------------------------------------------------------------------------

Let us remind briefly how to calculate this diagram by using the
nonrelativistic technics developed in \cite{Shapiro}. The
amplitude $M$ of this processus can be written as
\begin{eqnarray*}
%\label{slsquare}
M &=& \int \frac{d^3p_1dE_1}{(2\pi)^4}
\int \frac{d^3p_2dE_2}{(2\pi)^4} M_A \nonumber \\
&&\frac{-2im_{\tiny \mbox{p}}}
{\mbox{\bf p}_1^2-2m_{\tiny \mbox{p}}E_1-i0}\,
\frac{-2im_{\tiny \mbox{p}}}
{\mbox{\bf p}_2^2-2m_{\tiny \mbox{p}}E_2-i0}
\,M_{\tiny \mbox{d}}  \nonumber \\
&& (2\pi)^4 \delta^3 (\mbox{\bf p}_1+\mbox{\bf p}_2- \mbox{\bf P})
\delta (E_1+E_2- E-\varepsilon) \nonumber
\end{eqnarray*}
Here $M_A$ is the diagram corresponding to the block A
(production of nucleons 1 and 2 and other particles in the
final state $f$), $M_{\tiny \mbox{d}}$ the vertex of
coalescence to the deuteron, $m_{\tiny \mbox{p}}$ the nucleon mass.
Two fractions are propagators of nucleons 1 and 2, the integrals
are done over energies and momenta of these virtual particles.
The last delta functions reflect energy-momentum conservation in
the deuteron vertex ($\mbox{\bf P}= \mbox{\bf p}_1+\mbox{\bf p}_2$
is deuteron momentum, $E=\mbox{\bf P}^2/4m_{\tiny \mbox{p}}$
its kinetic energy, $\varepsilon \approx -2.2$~MeV its binding energy).

After trivial integration over $\mbox{\bf p}_2$ and $E_2$
and introduction of relative momentum
$\mbox{\bf q} =  \frac{1}{2}(\mbox{\bf p}_1-\mbox{\bf p}_2)$,
one obtains the following expression:
\begin{eqnarray*}
%\label{slsquare}
M &=& \int \frac{d^3q dE_1}{(2\pi)^4}
 M_A(\mbox{\bf P}, \mbox{\bf q}) M_{\tiny \mbox{d}}
 \frac{-2im_{\tiny \mbox{p}}}
{(\mbox{\bf q}+\mbox{\bf P}/2)^2-2m_{\tiny \mbox{p}}E_1-i0}\nonumber \\
&&
\frac{-2im_{\tiny \mbox{p}}}
{(\mbox{\bf q}-\mbox{\bf P}/2)^2-2m_{\tiny \mbox{p}}
(\mbox{\bf P}^2/4m_{\tiny \mbox{p}}-E_1+\varepsilon)-i0}
\end{eqnarray*}
The final integration over $dE_1$ (one supposes as usually that
the only singularities in the complex $E_1$ plane are those of propagators)
gives
\begin{eqnarray}
\label{Kolybamp}
M = i \int \frac{d^3q}{(2\pi)^3}
 M_A(\mbox{\bf P}, \mbox{\bf q}) \varphi_d(\mbox{\bf q})
\end{eqnarray}
where the deuteron wave function %$\varphi_d(\mbox{\bf q})$
\begin{eqnarray*}
\varphi_d(\mbox{\bf q}) =
 \frac{m_{\tiny \mbox{p}} M_{\tiny \mbox{d}}}
{\mbox{\bf q}^2+m_{\tiny \mbox{p}}|\varepsilon|}
\end{eqnarray*}
is normalized by the condition
\begin{eqnarray*}
\int \frac{d^3q}{(2\pi)^3} |\varphi_d(\mbox{\bf q})|^2=1.
\end{eqnarray*}

To perform further calculations
one needs to make an assumption about
the dependance of the amplitude $M_A$ corresponding to the block A
on its variables (the particle momenta). It can be shown
\cite{Kolyb}, that the simplest hypothesis that this amplitude is
constant gives rise to a wrong result: the production
cross-section appears to be zero. One can see it from (\ref{Kolybamp}).
When $M_A$ does not depend on momenta the transition amplitude $M$ becomes
proportional to
\begin{eqnarray*}
M = i M_A \int \frac{d^3q}{(2\pi)^3}
  \varphi_d(\mbox{\bf q}) \propto \varphi_d(\mbox{\bf r}=0)=0,
\end{eqnarray*}
i.e. the deuteron wave function at the origin which, for realistic
potentials, is equal to zero.

Therefore, the momentum
dependance of $M_A$ has to be introduced, for instance, in a ``minimal''
way: the inclusive nucleon spectra usually have a decreasing form
and can be parameterized by a Gaussian function in rather wide
parameter regions:
\begin{eqnarray}
E_{\tiny \mbox{p}}\frac{d^3\sigma_{\tiny \mbox{p}}}{dp_{\tiny
\mbox{p}}^3} \propto \exp \left( - \mbox{\bf p}_{\tiny
\mbox{p}}^2/Q^2 \right),
\end{eqnarray}
where $Q$ is related to the slope parameter. Accordingly, the
amplitude $M_A$ can be written in the following way:
\begin{eqnarray}
%\label{slsquare}
M_A =C \exp \left( - \frac{\mbox{\bf p}_1^2+\mbox{\bf
p}_2^2}{2Q^2} \right) =C \exp \left( - \frac{\mbox{\bf P}^2}{4Q^2}
-
\frac{\mbox{\bf q}^2}{Q^2} \right),
\end{eqnarray}
where the center-of-mass motion of the two nucleons is separated
from their relative motion. The amplitude $M_A$ determines the
cross-section for simultaneous production of two nucleons which
can be expressed in a standard way which supposes statistical
independence in production of the two nucleons as a product of
inclusive cross-sections:
\begin{eqnarray}
\frac{d^6\sigma_{\tiny \mbox{pn}}}{dp_{\tiny \mbox{p}}^3
dp_{\tiny \mbox{n}}^3} =
\frac{1}{\sigma_{inel}}\frac{d^3\sigma_{\tiny
\mbox{p}}}{dp_{\tiny \mbox{p}}^3} \frac{d^3\sigma_{\tiny
\mbox{n}}}{dp_{\tiny \mbox{n}}^3},
\end{eqnarray}
where $\sigma_{inel}$ is the cross-section of inelastic
interaction of initial particles.

After the substitution of (3) into the expression for the diagram
of Fig.~1 and taking into account (4), the cross-section for the
formation of deuterons takes the form
\begin{eqnarray}
E_{\tiny \mbox{d}}\frac{d^3\sigma_{\tiny \mbox{d}}}{dp_{\tiny
\mbox{d}}^3} = 12\pi^3 |S|^2 \frac{1}{m_{\tiny \mbox{p}}
\sigma_{inel}}E_{\tiny \mbox{p}} \frac{d^3\sigma_{\tiny
\mbox{p}}}{dp_{\tiny \mbox{p}}^3} E_{\tiny
\mbox{n}}\frac{d^3\sigma_{\tiny \mbox{n}}}{dp_{\tiny \mbox{n}}^3},
\end{eqnarray}
where $\mbox{\bf
p}_{\tiny \mbox{d}} \approx 2\mbox{\bf p}_{\tiny \mbox{p}}$, and
\begin{eqnarray}
S= \int \frac{d^3q}{(2\pi)^3} \varphi_d(\mbox{\bf q}) \exp \left(
- \frac{\mbox{\bf q}^2}{Q^2} \right).
\end{eqnarray}

The structure of (4) is exactly the same as that of the
coalescence model and one can obtain easily the following
expression of the coalescence radius in momentum
space\footnote{In this definition of $p_0$, the spin of outgoing
particles is not taken into account (see \cite{Kolyb} for
discussion). Note also that the definition of the parameter $p_0$
given in \cite{Salati} differs from the standard one by a factor
of 2.}
\begin{eqnarray}
p_0^3=  36\pi^2\left| \int \frac{d^3q}{(2\pi)^3}
\varphi_d(\mbox{\bf q}) \exp \left( - \frac{\mbox{\bf q}^2}{Q^2}
\right) \right|^2.
\end{eqnarray}

Thus, the approach based on the diagram of Fig.~1 reproduces the
coalescence model with $p_0$ which is no more a free parameter but
it is determined by the inclusive proton spectrum and by the
deuteron wave function. As an example, in Fig.~2 the values of
$p_0$ as a function of $Q$ (equation (8)) are presented for
different nucleon-nucleon potentials (Paris \cite{Paris}, Bonn
\cite{Bonn}, and different versions of Nijmegen potential
\cite{Nijmegen}).

%---------------------------------------------------------------------------
\begin{figure}[h!]
\epsfxsize=9cm
\centerline{\epsfbox{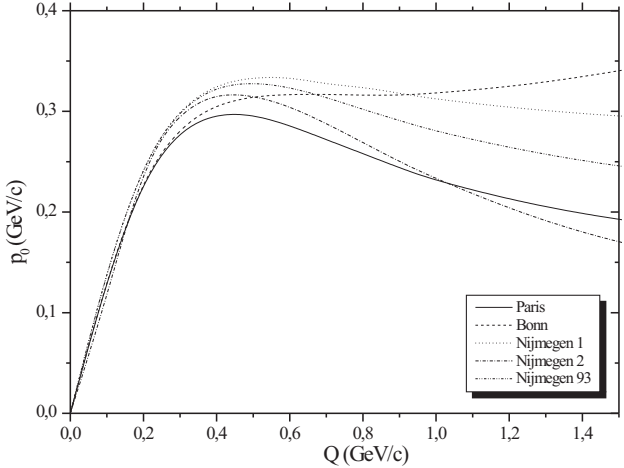}} \caption{ Dependence of the
coalescence momentum $p_0$ on the slope parameter $Q$ of inclusive
nucleon spectrum for different nucleon-nucleon potentials. }
\end{figure}
%
%---------------------------------------------------------------------------

One can see that all potentials give the same result up to $Q
\approx 300 $ MeV/$c$ where the deuteron wave function in momentum
space is quite well known. For higher values of $Q$ the difference
between the predictions of different nucleon-nucleon potentials
can be very important (taking into account the fact that the
cross-section is proportional to the third power of the
coalescence radius in momentum space $p_0$). In this work, the
Paris potential was chosen for further calculations.

The isotropic angular dependence supposed in (2) is quite
frequently used in nonrelativistic collisions. In the relativistic
case, the dependencies on transversal and longitudinal momentum
can be very different. However, the formulae obtained within this
approach can be easily generalized to any angular dependence.

If the inclusive nucleon cross-section is parameterized by an
amplitude $M_{\mbox{\bf p}_1}$
\begin{eqnarray}
E_1\frac{d^3\sigma_1}{dp_1^3} = |M_{\mbox{\bf p}_1}|^2,
\end{eqnarray}
the cross-section for the deuteron formation can be written as
(see (\ref{Kolybamp}))
\begin{eqnarray}
E_{\tiny \mbox{d}}\frac{d^3\sigma_{\tiny \mbox{d}}}{dp_{\tiny
\mbox{d}}^3} = \frac{12\pi^3}{\sigma_{inel} m_{\tiny \mbox{p}}}
\left|\int \frac{d^3q}{(2\pi)^3} M_{\mbox{\bf p}_1}M_{\mbox{\bf
p}_2}\varphi_d(\mbox{\bf q}) \right|^2.
\end{eqnarray}

It is clear that this model can be practically directly used to
describe the production of antideuterons. The only problem is a
presence of the threshold in the antiparticle production
cross-sections. The coalescence model and the approach used here
(10) are not valid in the near threshold region. Therefore one
needs to propose a phenomenological procedure to describe
experimental data near the threshold. For proton-proton
collisions, the authors of \cite{Salati} have proposed a quite
simple prescription: the center of mass energy available for the
production of the second antinucleon has to be reduced by twice
the energy carried away by the first antinucleon $E_{\tiny
\mbox{\=p}}$. In other words, the two antinucleons are supposed to
be produced at different energies: $\sqrt{s}$ and
$\sqrt{s}-2E_{\tiny \mbox{\=p}}$.

In this article, the antideuteron production threshold is taken
into account in a slightly different way. In proton-proton
collisions, the main reaction giving anti\-deuterons is $\mbox{pp}
\rightarrow \mbox{\=dpppn}$. Near the threshold of this reaction,
the energy dependence of the antideuteron production cross-section
is mostly determined by the phase space of four nucleons
$\Phi(\sqrt{s}-E_{\tiny \mbox{\=d}}; m, m, m, m)$,
\begin{eqnarray}
E_{\tiny \mbox{\=d}}\frac{d^3\sigma_{\tiny
\mbox{\=d}}}{dp_{\tiny \mbox{\=d}}^3} \propto
\Phi(\sqrt{s}-E_{\tiny \mbox{\=d}}; m, m, m, m).
\end{eqnarray}

The phase space $\Phi$ for $n$ particles with masses, momenta and energies,
respectively, $m_i$, $\mbox{\bf p}_i$, $E_i$ is defined in usual way
(in cms)
\begin{eqnarray*}
&&\Phi(\sqrt{s}; m_1, m_2, \ldots m_n) =\\
&&\prod_{i=1}^{n}  \frac{1}{(2\pi)^3} \frac{d^{3}p_i}{2E_i}
\delta^{3}\left(\sum_{i=1}^{n} \mbox{\bf p}_i\right)
\delta \left(\sum_{i=1}^{n} \mbox{\bf E}_i-\sqrt{s}\right)
\end{eqnarray*}
and is calculated by using standard CERN library program
\cite{CERNlib}. $\sqrt{s}$ is the total energy available for these
$n$ particles in the center of mass system.

Therefore one can introduce a phenomenological correction factor
$R$ to the formula (10) defined as
\begin{eqnarray}
R(\sqrt{s}-E_{\tiny \mbox{\=d}})=
\frac{\Phi(\sqrt{s}-E_{\tiny \mbox{\=d}}; m, m, m, m)}
{\Phi(\sqrt{s}-E_{\tiny \mbox{\=d}}; 0, 0, 0, 0)},
\end{eqnarray}
where the denominator contains the ultrarelativistic phase spa\-ce
to ensure $R$ to be dimensionless and to have correct behavior at
high energies ($R \rightarrow 1$). The behavior of $R(x)$ is
presented in Fig.~3.
%---------------------------------------------------------------------------
\begin{figure}[h!]
\epsfxsize=9cm
\centerline{\epsfbox{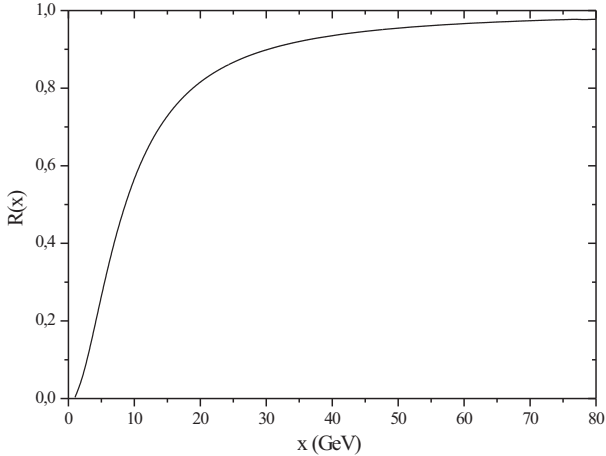}} \caption{ Dependence of the
threshold factor $R(x)$ on its argument. }
\end{figure}
%
%---------------------------------------------------------------------------

There are, at least, two
advantages with respect to the prescription chosen in
\cite{Salati}. Firstly, one makes no assumption about mechanism of
the production. Secondly, this correction factor has correct
kinematic behavior both near the threshold and at high energies.
We'll discuss the role of this factor later.

%%%%%%%%%%%%%%%%%%%%%%%%%%%%%%%%%%%%%%%%%%%%%%%%%%%%%

\section{Description of experimental data}

Before presenting of the results let us make some preliminary
remarks about existing experimental data.
\begin{itemize}

\item The experimental data on antideuteron production are not abundant
and are much less informative than that on deuteron production.
There are only a few different experimental observations of
antideuteron production in proton-proton
\cite{ISRdbar73,ISRdbar78,ISRdbar75,ISRout,IHEPpp},
proton-nuc\-leus \cite{IHEPpp,IHEPpAl69,IHEPpAl71,FL}, and
nucleus-nucleus collisions \cite{Aoki,AGSAuPb,NA52}. Some
experimental results cannot be analyzed within our approach
because the experimental information is not complete. For
instance, in \cite{Dorfan} there is no data on antiproton
production for corresponding energies; in \cite{Bozzoli} only
relative spectra (antiprotons to $\pi^-$ and antideuterons to
$\pi^-$) were measured.

\item To obtain a reasonable description within the present approach
one needs to know the cross-section of antiproton production for
the antiproton momentum equal to a half of the antideuteron one.
Unfortunately, this condition is rarely satisfied: in most
experiments, the differential cross-sections of the antiproton and
antideuteron production being measured for approximately the same
momentum. Therefore, to apply the method one has to extrapolate
the antiproton data to another kinematical region. This procedure,
of course, introduces an additional error.

\item In principle, the inclusive cross-sections discussed here are
functions of two kinematical variables (for instance, transversal
and longitudinal momentum) and one has to present the results in
three-dimensional form. However, in each experiment one has only a
few experimental points and the results are presented as a
function of one variable (either total or transversal momentum).

\end{itemize}

The total inelastic cross-section was taken from the PDG data
\cite{PDG} (for proton-proton collisions) or described by the
well-known parameterization \cite{Sigmainel} (for proton-nucleus
and nucleus-nucleus collisions).

\subsection{Proton-proton collisions}

Let us begin the analysis with the most informative experiment
performed on the ISR at CERN. In a few experi\-ments, the spectra
of antiprotons \cite{ISRpbar} and antideuterons
\cite{ISRdbar73,ISRdbar78} were measured in pp-collisions at
$\sqrt{s}=53$~GeV. The detector was situated at 90 degrees (in
this geometry, the total momentum of outgoing particles coincides
with the transversal one). Inclusive antiproton cross-sections
over quite large regions of momentum and $\sqrt{s}$ was obtained
in \cite{ISRpbar}. An example of an experimental distribution of
antiprotons for $\sqrt{s}=53$~GeV is presented in Fig.~4 in
comparison with two different parameterizations of the data.

%---------------------------------------------------------------------------
\begin{figure}[h!]
\epsfxsize=9cm \centerline{\epsfbox{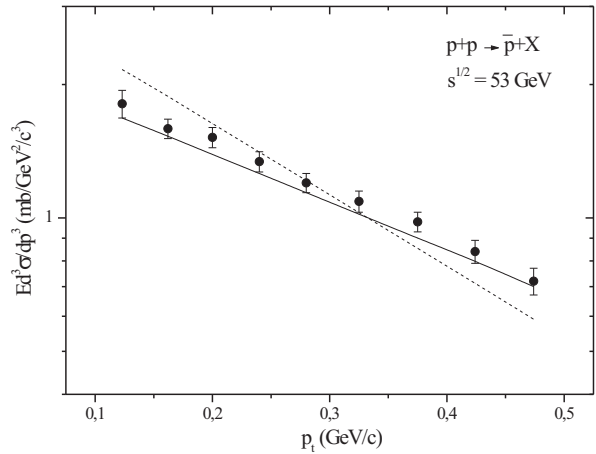}} \caption{The
inclusive differential cross-section of antiproton production as a
function of the transversal momentum $p_t$ compared to two
different parameterizations: solid line -- exponential one
\cite{ISRpbar}; dashed line -- Tan and Ng \cite{TanNg}
parameterization. The data are taken from \cite{ISRpbar}.}
\end{figure}
%
%---------------------------------------------------------------------------

The first parameterization (solid line) proposed by the authors of
the experiment \cite{ISRpbar}:
\begin{eqnarray}
E_{\tiny \mbox{\=p}} \frac{d^3\sigma_{\tiny
\mbox{\=p}}}{dp_{\tiny \mbox{\=p}}^3} = As^\alpha \exp[-Bp_t],
\end{eqnarray}
with parameters $A=0.195$~mbarn/(GeV/$c$)$^2$, $\alpha = 0.310$,
$B=2.49$~(GeV/$c$)$^{-1}$ describes perfectly these data. The
second one (dashed line)
\begin{eqnarray}
E_{\tiny \mbox{\=p}} \frac{d^3\sigma_{\tiny
\mbox{\=p}}}{dp_{\tiny \mbox{\=p}}^3} = f\exp[-(Ap_t+Bp_t^2)],
\end{eqnarray}
is the frequently used parameterization proposed by Tan and Ng
\cite{TanNg} which works quite well in a wide region of $p_t$ and
$\sqrt{s}$. Here $f=f(E^*, \sqrt{s})$, $A=A(E^*, \sqrt{s})$ and
$B=B(E^*, \sqrt{s})$ are known functions of $\sqrt{s}$  and of the
antiproton energy in center-of-mass system $E^*$. This formula
gives here reasonable values of the cross-section but the trend is
not well reproduced. We present here both quite close
parameterizations to demonstrate the difference in description of
the data on antideuteron production. The corresponding
cross-section measured in this experiment
\cite{ISRdbar73,ISRdbar78} are given in Fig.~5 and compared with
different calculations.

%---------------------------------------------------------------------------
\begin{figure}[h!]
\epsfxsize=9cm \centerline{\epsfbox{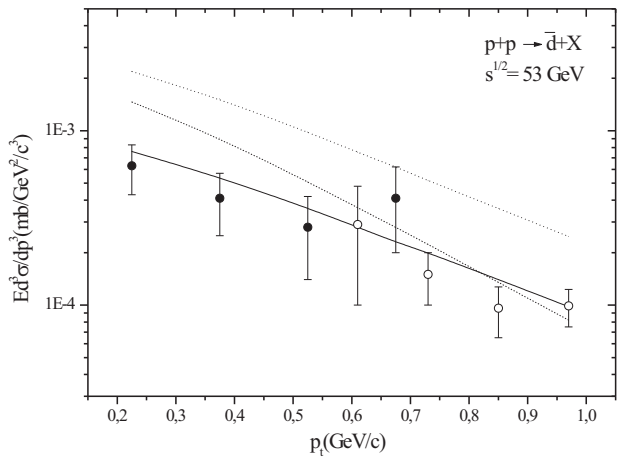}} \caption{The
inclusive differential cross-section of antideuteron production as
a function of the transversal momentum $p_t$ compared to the
calculations with two different parameterizations of the
antiproton production cross-section: solid line -- exponential
fit; dashed line -- Tan and Ng parameterization. Dotted line is
the calculation with an exponential function without the
anisotropy effect. The data are taken from \cite{ISRdbar73} (black
circles) and \cite{ISRdbar78} (open circles).}
\end{figure}
%
%---------------------------------------------------------------------------

Comparison between the two parameterizations shows that the
description of the antideuteron production is quite sensitive to
the antiproton production cross-sections: 20-30$\%$ difference in
description of the \={p} data can result in a factor of 2 for the
\={d}. This difference can be even more pronounced if one has to
extrapolate a chosen parameterization. As we mentioned previously
$p_{\tiny \mbox{\=d}} \approx 2p_{\tiny \mbox{\=p}}$. This
condition is satisfied for these ISR data. In some other
experiments presented hereafter, it is not the case.

In this figure, one can see also importance of the aniso\-tropy of
angular distributions. An exponential parameterization (13) of the
antiproton production cross-section can be seen in two ways: as a
function of the total antiproton momentum or of the transversal
one (as stated previously, in the particular geometry of this
experiment, they are equal to each other). However in the integral
(10), all directions (and not only transverse one) are presented
and the difference between total and transversal momenta can be
quite important. Thus for the data under consideration the solid
line represents the results with parameterization (13) whereas the
dotted line corresponds to the same parameterization but with
total momentum instead of trans\-versal one (antiproton production
cross-section supposed to be isotropic).

By using the picture of coalescence model, it is quite easy to see
when anisotropy can be important in description of experimental
data. If the total antideuteron momentum $p_{cm}$ is very high
with respect to the coalescence radius in momentum space $p_0$,
antiproton and antineutron are produced in approximately the same
direction (the direction of antideuteron momentum) and an
anisotropy of antinucleon angular distributions plays no role. If
$p_0\approx p_{cm}$, the two antinucleons can propagate in quite
different directions before the coalescence and it is necessary to
take anisotropy into account correctly.

For these ISR data, the total antideuteron momentum $p_{cm}$ is of
the order of the coalescence momentum $p_0$ and the anisotropy
effect is seen clearly. In all other experimental data discussed
hereafter, $p_{cm}$ is very high with respect to $p_0$ and the
anisotropy effect is not so important.

On can thus understand easily that if one uses aniso\-tropic
cross-sections the effect of the D-wave in the deuteron wave
function can be quite important (for isotropic Gaussian
para\-metrization (1), the D-wave contribution is explicitly equal
to 0). For these ISR data, the introduction of D-wave contribution
into the deuteron wave function divides the value of the
cross-section by a factor of 2.

In all calculations presented in this article, the Paris wave
function is used. For most experimental data analyzed in this
article, the choice of the deuteron wave function is not crucial:
if characteristic slope parameter of the inclusive antinucleon
spectrum is less than approximately 0.5 GeV/$c$, all potential
models give close values of $p_0$ (see Fig.~2). However, for some
sets of the data it is not the case (we will mention it where
necessary).

%Let us emphasize once more that these calculations contain no
%free parameter.

Another set of experimental data for the same $\sqrt{s}$ but with
very large longitudinal component of antideuteron momentum $p_l
\approx 5-7 $ GeV/$c$ was measured in \cite{ISRdbar75} and is
presented in Fig.~6. The antiproton spectrum is taken from an
experiment performed by this group \cite{pbarforISR75}.
Unfortunately, the parameterization of Tang and Ng does not work
well here (its prediction exceeds systematically the data by a
factor of 2). Therefore, we fitted the data by a Gaussian
(proposed also by the authors of
 \cite{pbarforISR75}) and by an exponential function of $p_t$. Both
parameterizations give quite good antideuteron production cross
section.

%---------------------------------------------------------------------------
\begin{figure}[h!]
\epsfxsize=9cm \centerline{\epsfbox{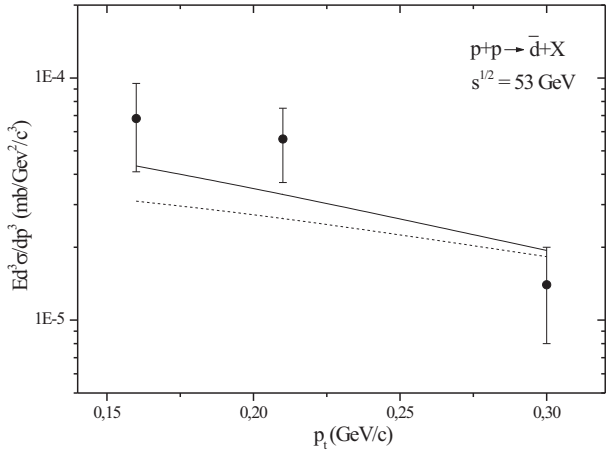}} \caption{The
inclusive differential cross-section of antideuteron production as
a function of the transversal momentum $p_t$ compared to the
calculations with two different parameterizations of the
antiproton production cross-section: solid line -- exponential
function; dashed line -- Gaussian. The data are taken from
\cite{ISRdbar75}.}
\end{figure}
%
%---------------------------------------------------------------------------

However, it appears to be impossible to describe reasonably
another set of ISR data measured at very big transfer momentum
\cite{ISRout}. The parameterization of Tang and Ng cannot
reproduce the antiproton production cross-section: the order of
magnitude is correct but not a general trend which has quite
unusual form (the cross-section increases with $p_t$ increasing!
Thus the absence of the \={p} cross-section parameterization does
not allow us to obtain a  reasonable description of the \={d}
production data (the use of Tan and Ng parameterization as input
gives a result 4--10 times higher than the experimental data).

The last measurement of \={d} production in pp-collisions was
performed ten years later at the IHEP machine \cite{IHEPpp} at
lower cms energies ($\sqrt{s}=11,5$~GeV) and with different
geometry (with fixed target). The data were taken at very high
$p_t$ and $p_l$ and are not very rich (two points both for \={p}
and \={d}). Tang and Ng parameterization fails to describe the
data (by a factor of ten for the highest momentum) and the
characteristic slope parameter of the inclusive \={p} spectrum is
quite high (where different potential models give quite different
(by a factor of 2) predictions). However, we decided to present
this instructive example because one can estimate here the role of
the threshold effect by the procedure proposed in (12). One can
fit the antiproton data (two points) both by a Gaussian and an
exponential function of $p_t$. The corresponding predictions for
the \={d} production cross-section are given in Fig.~7 by dotted
and solid line respectively (without threshold effect). Once the
threshold is taken into account by introduction of the factor, $R$
(12), the agreement with the experimental data is improved
significantly (dashed line represents the calculations with an
exponential parameterization and the threshold effect).

%---------------------------------------------------------------------------
\begin{figure}[h!]
\epsfxsize=9cm \centerline{\epsfbox{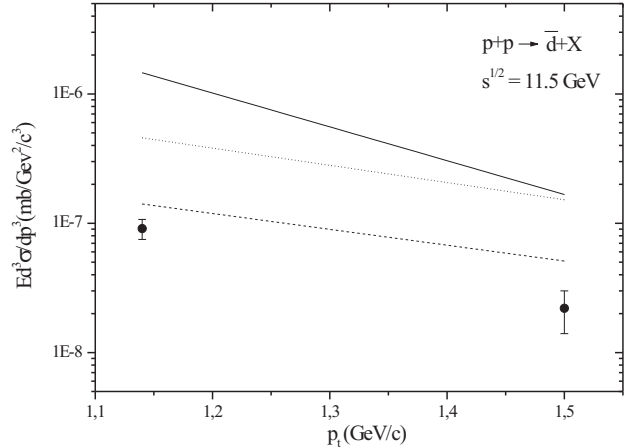}} \caption{The
inclusive differential cross-section of antideuteron production as
a function of the transversal momentum $p_t$ compared to the
calculations with two different parameterizations of the
antiproton production cross-section: solid line -- exponential
function; dotted line -- Gaussian. In both calculations, the
threshold effect is not included. Third line (dashed) corresponds
to calculations using the Gaussian parameterization including
the threshold effect. The data are taken from \cite{IHEPpp}.}
\end{figure}
%
%---------------------------------------------------------------------------

\subsection{Proton-nucleus collisions}

Unfortunately, there are no more exploitable data on the
antideuteron production in proton-proton collisions. To test
further the model, we analyzed available data on the \={d}
production in proton-nucleus collisions. Here, one has no more
general parameterization like \cite{TanNg} and, for each set of
data, a different parameterization is used \footnote{Note that in
high energy proton-nucleus and nucleus-nucleus collisions, the
physical center-of-mass system is that of nucleon (from the
target) - nucleon (from the beam) one.}.

In the same IHEP experiment \cite{IHEPpp}, the production of
antideuterons was measured on Be and Pb targets also.

For these data, one can make the same remarks as for the data
obtained in the proton-proton collisions (only a few experimental
points, very high momenta of outgoing particles, and quite big
value of the slope parameter). The results are presented in Fig.~8
and look quite encouraging.

%---------------------------------------------------------------------------
\begin{figure}[h!]
\epsfxsize=9cm \centerline{\epsfbox{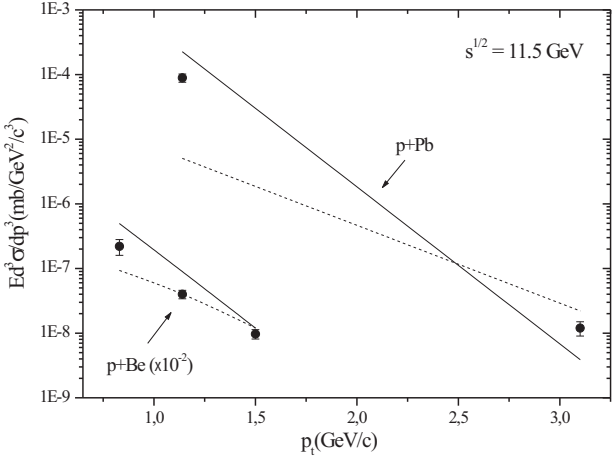}} \caption{The
inclusive differential cross-section of antideuteron production on
Be and Pb targets as a function of the transversal momentum $p_t$
compared to the calculations with two different parameterizations
of the antiproton production cross-section: solid line --
exponential function; dashed line -- Gaussian. The data are taken
from \cite{IHEPpp}.}
\end{figure}
%
%---------------------------------------------------------------------------

There are also old IHEP measurements of the anti\-deuteron
production in p-Al collisions \cite{IHEPpAl69,IHEPpAl71}. The data
were taken in the forward (or practically forward) direction and
we have no information about $p_t$ dependence of the antiproton
production cross-section (thus we can not take completely into
account the anisotropy of angular distributions). Therefore, in
Fig.~9, the results of calculations and the data are presented as
a function of the total \={d} momentum. Note that, for the
Gaussian function, the value of the slope parameter $Q$ appears to
be very large (of the order of 1.3~GeV/$c$) where the deuteron
wave function is not known and the results depend strongly on the
nuclear potential.

%---------------------------------------------------------------------------
\begin{figure}[h!]
\epsfxsize=9cm \centerline{\epsfbox{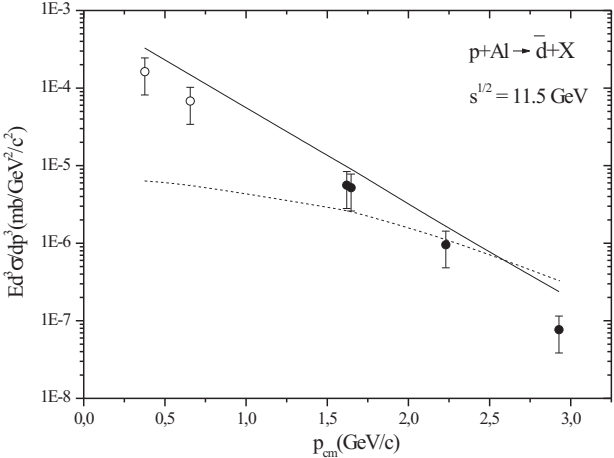}} \caption{The
inclusive differential cross-section of antideuteron production on
Al target as a function of the total momentum in the center-of
mass frame compared to the calculations with two different
parameterizations of the antiproton production cross-section:
solid line -- exponential function; dashed line -- Gaussian. The
data are taken from \cite{IHEPpAl69} (black circles) and
\cite{IHEPpAl71} (open circles).}
\end{figure}
%
%---------------------------------------------------------------------------

This analysis can be completed by the measurement performed in
FNAL \cite{FL} where the antideuteron production cross-section was
measured for different targets (Be, Ti, W) at intermediate (with
respect to the ISR and IHEP experiments) energies ($\sqrt{s} =
23.7$~GeV/$c$). The data were taken at quite high transverse
momentum. The theoretical results presented in Fig.~10 are in quite
good agreement with the experimental data for the three targets.
One of the reasons for this good agreement is a good knowledge of
the antiproton production cross-section obtained in this
experiment.

%---------------------------------------------------------------------------
\begin{figure}[h!]
\epsfxsize=9cm \centerline{\epsfbox{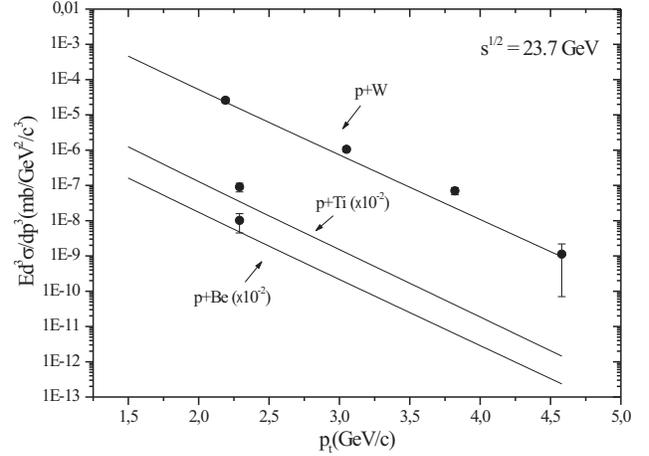}} \caption{The
inclusive differential cross-section of antideuteron production on
Be, Ti, and W targets as a function of the transversal momentum
$p_t$ compared to the calculations with exponential
parameterization of the antiproton production cross-section. The
data are taken from \cite{FL}.}
\end{figure}
%
%---------------------------------------------------------------------------

It is necessary to note that there are also some experimental
results on antideuteron production in nucleus-nucleus collision
obtained in the AGS experiment (one point in Si+Al collision
\cite{Aoki} and two points in Au+Pb \cite{AGSAuPb}) and in the
NA52 experiment \cite{NA52}. Unfortunately here, one has neither a
good parameterization of the antiproton production cross-section
nor a reliable parameterization for the total inelastic
cross-section. The measurements were made in the forward direction
and one has no information about the $p_t$ dependance of the
antiproton production cross-sections (thus it is impossible to
take into account correctly the anisotropy of angular
distributions which can be very important here). Simple
parameterizations (Gaussian and exponential) gave no satisfactory
description. In general, the discrepancy between the calculations
and the data is of order of a factor 5, which is not very
surprising taking into account all these remarks. As an example,
in Fig.~11 the calculation of the antideuteron production cross
section in Pb+Pb collisions as a function of total momentum in the
center-of-mass frame in comparison with the experimental data
\cite{NA52} are presented. The data cover a very large momentum
region and the characteristic momentum in the antiproton
production cross-section is very big (of the order of 2 GeV/$c$).

%---------------------------------------------------------------------------
\begin{figure}[h!]
\epsfxsize=9cm \centerline{\epsfbox{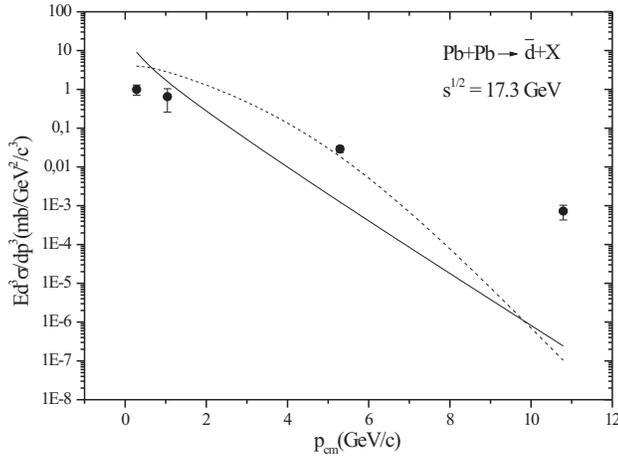}} \caption{The
inclusive differential cross-section of antideuteron production in
Pb-Pb collisions as a function of total momentum in the center of
mass frame compared to the calculations with two different
parameterizations of the antiproton production cross-section:
solid line -- exponential function; dashed line -- Gaussian. The
data are taken from \cite{NA52}.}
\end{figure}
%
%---------------------------------------------------------------------------

\section{Conclusions}

Our main conclusion is that the diagrammatic approach to the
coalescence model developed in \cite{Kolyb} can be successfully
applied to the description of the antideuteron production in
proton-proton and proton-nucleus collisions. There are two
modifications: firstly, it is necessary to take into account the
threshold effect and, secondly, one must include specific
consideration of the strong anisotropy of angular distributions of
antiproton production. Once these phenomena are taken into
account, the model can describe most of existing
experimental data on antideuteron production in proton-proton and
proton-nucleus collisions.

The successful reproduction of experimental data suggests that
good knowledge of the antinucleon production cross-section and of
the deuteron wave function allow to describe the antideuteron
production cross-section in quite large region of kinematic
variables without any additional parameter.

\section*{Acknowledgments}

During the preparation of this article, we learnt the death of our
friend and colleague V.M.~Kolybasov who proposed the model
discussed in this article. We would like to dedicate this work to
his memory.

The authors would like to thank A.~Barrau, M.~Bu\'enerd, A.J.~Cole,
L.~D\'erome, V.A.~Karmanov and P.~Salati for very useful and simulating
discussions.

\end{document}